# ARTICLE

# Ni-Sn intermetallics as efficient buffering matrix of Si anodes in Li-ion batteries

Tahar Azib,[a] Nicolas Bibent,[b] Michel Latroche,[a] Florent Fischer,[c] Jean-Claude Jumas,[b] Josette Olivier-Fourcade,[b] Christian Jordy,[c] Pierre-Emmanuel Lippens[b] and Fermin Cuevas[a]

For a successful integration of silicon in high-capacity anodes of Li-ion batteries, its intrinsic capacity decay on cycling due to severe volume swelling should be minimized. In this work, Ni-Sn intermetallics are studied as buffering matrix during reversible lithiation of Si-based anodes. Si/Ni-Sn composites have been synthetized by mechanical milling using C and Al as process control agents. $Ni_3Sn_4$, $Ni_3Sn_2$ intermetallics and their bi-phasic mixture were used as constituents of the buffering matrix. The structure, composition and morphology of the composites have been analyzed by X-ray diffraction (XRD), $^{119}Sn$ Transmission Mössbauer Spectroscopy (TMS) and scanning electron microscopy (SEM). They consist of ~ 150 nm Si nanoparticles embedded in a multi-phase matrix, the nanostructuration of which improves on increasing the $Ni_3Sn_4$ amount. The electrochemical properties of the composites were analyzed by galvanostatic cycling in half-cells. Best results for practical applications are found for the bi-phasic matrix $Ni_3Sn_4$-$Ni_3Sn_2$ in which $Ni_3Sn_4$ is electrochemically active while $Ni_3Sn_2$ is inactive. Low capacity loss, 0.04 %/cycle, and high coulombic efficiency, 99.6%, were obtained over 200 cycles while maintaining a high reversible capacity above 500 mAh/g at moderate regime C/5.

## Introduction

Energy storage demands for current and future portable electronic devices, electrical vehicles and management of renewable energies are ever growing [1–3]. The performances of Li-ion batteries and related technologies are expected to face these demands [4–6]. The development of advanced anode materials for lithium ion batteries exhibiting higher capacity than current graphite electrodes, $C = 370$ mAh $g^{-1}$, remains critical [7]. This can be potentially achieved by alternative reaction schemes to classical intercalation such as conversion or alloying reactions [8,9]. Among the latter, silicon-based anode materials have attracted much interest as negative electrodes of Li-ion batteries due to the high capacity of Si, $C = 3600$ mAh $g^{-1}$, low potential and environmental friendliness [10–12]. Nevertheless, silicon suffers from low electronic conductivity, unstable solid electrolyte interface, and severe pulverization triggered by large volume changes (~300%) upon lithium alloying and extraction [11,13–15]. This causes low coulombic efficiency and rapid capacity fading. Recently, many research efforts have been made to overcome this issue by modification of size, shape, and crystallinity of silicon. Original microstructures from 1D-nanowires to 2D thin films have been synthesized [16–21]. For high-energy storage applications, bulk 3D microstructures are more suitable [22–24]. Embedding silicon in a matrix that helps to accommodate volume changes and improve electronic conduction is expected to overcome the limitations of Si as anode material.

Among different matrix materials, embedding silicon in intermetallic compounds has received particular attention, especially in the industrial sector [14,25,26]. Indeed, synthesis routes currently used to produce them, such as melting or solid-state mechanical milling, [24,27–29] are largely available in the metallurgical industry. These methods have the advantage, compared to some exotic processes, to produce materials at large scale and at reasonable cost. In the composite silicon/intermetallic approach, the intermetallic compound does not only accommodate the Si volume expansion but also improves the electronic conductivity of the composite material. Intermetallics can be formed by elements active and/or inactive toward lithium and therefore having different properties as buffering matrix of Si-anodes [30]. Among different intermetallics, Ni-Sn have attracted particular attention due to their high theoretical capacities and excellent cycling properties [31–37].

In this context, our group has developed in the recent years composite materials formed by Si nanoparticles embedded in a buffering matrix containing the intermetallic compound $Ni_3Sn_4$, Al and C [25,38,39]. Carbon acts as a process control agent (PCA) during mechanochemical synthesis of the composite by limiting the chemical reaction between Si and $Ni_3Sn_4$. A minor amount of aluminum (3 wt.%) is added since it is reported to improve the cycle-life of Si-Sn anodes [40]. Si/$Ni_3Sn_4$/Al/C composites provide high reversible capacity (above 700 mAh/g over 200 cycles) at low kinetic rates (C/50) but capacity loss (0.14 %/cycle) is still too high for long-standing applications. Moreover, material costs, driven by its high Sn content (46 wt.%), should be reduced. Indeed, the price of tin has

a. Univ Paris Est Creteil, CNRS, ICMPE, 2 rue Henri Dunant, 94320 Thiais, France.
b. ICGM, UMR 5253 CNRS, Université de Montpellier, Place Eugène Bataillon, 34095 Montpellier Cedex 5, France.
c. SAFT Batteries. 113 Bd. Alfred Daney, 33074 Bordeaux Cedex. France.

been multiplied by a factor of four in the last 20 years [41]. To mitigate these drawbacks, we here investigate the partial or full replacement of $Ni_3Sn_4$ by Sn-poorer $Ni_3Sn_2$. These two intermetallics are neighboring phases in the Ni-Sn phase diagram [42]. From this research, we demonstrate that bi-phasic $Ni_3Sn_4/Ni_3Sn_2$ matrix formed by electrochemically active/inactive counterparts, respectively, is highly beneficial. The bi-phasic matrix not only helps to decrease material costs but also to improve electrochemical performances as concerns long-term coulombic efficiency and cycle-life.

## Experimental

Single phase $Ni_3Sn_2$ and $Ni_3Sn_4$, and biphasic $Ni_3Sn_2$-$Ni_3Sn_4$ intermetallics (labelled as samples A, B and AB, respectively) were prepared by powder metallurgy from pure elements, Ni (99.9%, Sp < 45 μm, Cerac) and Sn (99.9%, Sp < 45 μm, Alfa-Aesar). For $Ni_3Sn_2$, its phase field $Ni_{3+z}Sn_2$ extends from $-0.30 \leq z \leq 0.22$ [42] and the stoichiometric compound $Ni_3Sn_2$ ($z = 0$) was targeted for the synthesis. For phase $Ni_3Sn_4$, the homogeneity range $Ni_{3+\varepsilon}Sn_4$ extends from $0.08 \leq \varepsilon \leq 0.6$ [43] and the Ni-rich compound $Ni_{3.6}Sn_4$ ($\varepsilon = 0.6$) was selected. The biphasic intermetallic AB was a mixture of nominal $Ni_3Sn_2$ and $Ni_{3.6}Sn_4$ compounds in ideal 1:1 weight ratio.

The powders were weighted according to the nominal intermetallic compositions, mechanically pelletized and sintered under argon atmosphere in a sealed silica tube. Annealing conditions were selected with respect to Ni-Sn phase diagram characteristics [42]. $Ni_3Sn_2$ compound (LT to HT polymorphic transition at 508 °C) was first annealed at 550 °C for 7 days and then at 400 °C for 7 days. $Ni_{3.6}Sn_4$ compound (peritectic reaction at 794 °C) was annealed at 700 °C for 7 days. The intermetallic mixture AB ($Ni_3Sn_2+Ni_{3.6}Sn_4$) was synthetized as for $Ni_3Sn_2$ single compound but the last thermal treatment was performed at 350 °C. After intermetallic synthesis, the pellets were mechanically pulverized and sieved down to 36 μm. Their chemical composition was analyzed by electron probe microanalysis (EPMA) using a Cameca SX-100 instrument, after embedding the powders in epoxy resin.

Three composites labeled as "Si-A", "Si-B" and "Si-AB" were prepared using the previously synthetized precursors $Ni_3Sn_2$ (A), $Ni_{3.6}Sn_4$ (B) and $Ni_3Sn_2$-$Ni_{3.6}Sn_4$ (AB), respectively. The composites were obtained by ball milling of the Ni-Sn intermetallics, silicon (99.9%, Sp < 1 μm), aluminum (99%, Sp ≤ 75 μm, Aldrich) and graphite. The weight percent of the starting milled powders is 65:19:3:13, respectively. Mechanical milling was performed under argon atmosphere in a Fritsch P7 planetary mill at a rotation speed of 600 rpm for 20 h. The ball-to-powder ratio was 5:1.

The crystal structure of the intermetallic precursors and the milled composites was examined by X-ray diffraction (XRD) with a Bruker D8 diffractometer using Cu $K_\alpha$ radiation in the 2θ range from 20° to 70°. Diffraction patterns were analyzed by the Rietveld method using the FullProf software [44]. The morphology and phase distribution of the composites were observed using a SEM-FEG MERLIN ZEISS scanning electron microscope (SEM) operated in back-scattered electron mode and equipped with Energy-Dispersive X-ray (EDX) analyzer. Complementary morphological and chemical analysis were done by transmission electron microscopy (TEM) with a Tecnai FEI F20 ST microscope providing high spatial resolution imaging as well as phase identification by Electron Energy Filtered Transmission

Electron Microscopy (EFTEM). Additional characterizations of the tin-based phases were performed with $^{119}Sn$ Transmission Mössbauer Spectroscopy (TMS). The spectra were recorded in transmission geometry and a constant acceleration mode at room temperature with a $Ca^{119m}SnO_3$ γ-ray source. The values of isomer shift (δ), quadrupole splitting (Δ), full line width at half-maximum (Γ) and relative spectral contribution (RSC) were determined by fitting Lorentzian lines to the experimental data with a nonlinear least-square method. The values of the isomer shift are given relative to $BaSnO_3$.

Electrochemical tests were carried out using 2032 coin-cells assembled under argon atmosphere. The working electrode material was prepared by mixing 55 wt.% of the 20 h-milled composite sieved under 125 μm, 20 wt.% of carboxymethyl-cellulose (CMC) as binder and 25 wt.% of carbon black. Such low loading of the composite material in the working electrode (55 wt.%) was expressly chosen to avoid limitations on its intrinsic properties due to electrode formulation. The slurries of the mixture were spread on Cu foil collector. The electrolyte was a 1:1:3 vol. mixture of ethylene carbonate (EC), propylene carbonate (PC) and dimethylcarbonate (DMC) containing 1 mol.$L^{-1}$ of $LiPF_6$. Lithium metal was used as negative electrode. The cells were cycled in galvanostatic mode using a Biologic VMP3 device. To promote electrode activation, first lithiation (discharge) was performed at a kinetic regime of C/50 down to 0 V vs. $Li^+/Li$ and second one at C/20 regime down to 70 mV vs. $Li^+/Li$ to avoid the formation of the crystalline $Li_{15}Si_4$ phase [45,46]. Charging steps (delithiation) were conducted at C/20 regime up to 2 V vs. $Li^+/Li$ for the first two cycles. Next cycles were performed at C/5 regime, both for discharge and charge steps, in the voltage range 0.07–2 V vs. $Li^+/Li$. At the end of discharge, constant voltage (CV) at 0.07 V was applied until a limiting current of 0.02 C. Rate capability studies were performed on charging. After 5 activation cycles, slow discharging of the electrode was done (C/70) to ensure full material lithiation. Next, sequential charging steps ranging from 20 C to 0,2C (i.e. C/5) up to a cut-off potential of 2V were applied, imposing 0.5 h relaxation between two consecutive charge steps. All electrochemical characterizations were performed at room temperature.

## Results

### Chemical and microstructural characterization

**Intermetallic Ni-Sn precursors.** The chemical composition of the three samples A, B and AB was analyzed by EPMA (see Fig. S1 and Table S1 in Supplementary Information (SI)). The composition of the major phases concurs with the targeted values, though minor stoichiometric deviations on Ni content are observed as shown in Table 1. Samples A and B are Ni-richer as compared to stoichiometric $Ni_3Sn_2$ and $Ni_3Sn_4$. For sample AB, $Ni_3Sn_2$ is Ni-poor ($Ni_{2.6}Sn_2$) whereas $Ni_3Sn_4$ is Ni-rich ($Ni_{3.7}Sn_4$). This is in good agreement with the reported binary Ni-Sn phase diagram [42] for which the limiting phase composition of the two-phase region $Ni_3Sn_2$-$Ni_3Sn_4$ is $Ni_{2.7}Sn_2$ and $Ni_{3.6}Sn_4$. It is also consistent with the significant homogeneity ranges reported for these two phases in the diagram.

The Rietveld analysis of XRD patterns of samples A, B and AB are shown in Fig. S2 and structural data are gathered in Table 1. All patterns display sharp diffraction peaks which evidences that intermetallic phases are well crystallized. For sample A, all diffraction peaks can be indexed in the orthorhombic LT modification of $Ni_3Sn_2$ (S.G. n°62, *Pnma*) with lattice parameters in agreement with those of the stoichiometric compound [47,48]. Refinement of Ni occupancy at

Wyckoff site 4*c* (*x*, ¼, *z*) led to the composition $Ni_{3.0(5)}Sn_2$. The diffraction pattern of sample B can be fully indexed in the monoclinic $Ni_3Sn_4$ phase (S.G. n°12, *C2/m*) with lattice parameters corresponding to Ni-rich $Ni_{3+ε}Sn_4$ compound with $ε$ ~ 0.6.

**Table 1.** Structural data of the A ($Ni_3Sn_2$), B ($Ni_3Sn_4$) and AB ($Ni_3Sn_2$-$Ni_3Sn_4$) samples. Standard deviations referred to the last digit are given in parenthesis.

| Sample | Phases | Composition (EPMA) | *S.G* | Composition (XRD) | Unit-cell parameters | | | | |
|---|---|---|---|---|---|---|---|---|---|
| | | | | | *a*, Å | *b*, Å | *c*, Å | β (°) | *V*, Å³ |
| A | $Ni_3Sn_2$ | $Ni_{3.08(3)}Sn_2$ | *Pnma* | $Ni_{3.0(5)}Sn_2$ | 7.121(1) | 5.193(1) | 8.145(1) | - | 301.2(1) |
| AB | $Ni_3Sn_2$ | $Ni_{2.59(1)}Sn_2$ | *Pnma* | $Ni_{2.6(1)}Sn_2$ | 7.000(3) | 5.120(2) | 8.078(3) | - | 289.9(1) |
| | $Ni_3Sn_4$ | $Ni_{3.69(2)}Sn_4$ | *C2/m* | $Ni_{3.6(1)}Sn_4$ | 12.48(2) | 4.085(2) | 5.206(2) | 103.37(4) | 258.6(1) |
| B | $Ni_3Sn_4$ | $Ni_{3.55(1)}Sn_4$ | *C2/m* | $Ni_{3.6(1)}Sn_4$ | 12.43(1) | 4.078(1) | 5.211(1) | 103.70(2) | 256.6(1) |

Here, refinement of Ni occupancy at Wyckoff site 2*a* (0, 0, 0) led to the composition $Ni_{3.6(1)}Sn_4$ [43]. As for sample AB, diffraction peaks can be indexed according to a combination of the two previous LT-$Ni_3Sn_2$ and $Ni_3Sn_4$ phases (with phase amount 51 and 49 wt.%, respectively), though for the former phase the unit-cell volume is significantly shrunk, $V$ = 289.9 Å³, as compared to that of the stoichiometric A sample, $V$ = 301.2 Å³. Such diminution evidences a lowering in Ni content as reported by Fjellvag and Kjekshus [47]. Indeed, the contribution of the $Ni_3Sn_2$ phase to the AB pattern is slightly better refined (not shown) in the incommensurate LT' modification that holds for Ni-contents lower than $Ni_{2.8}Sn_2$ [48].

The Mössbauer spectra of samples A (Fig. S3a) and B (Fig. S3b) were fitted with two doublets in agreement with the existence of two Sn crystallographic sites in both $Ni_3Sn_2$ and $Ni_3Sn_4$. The values of the Mössbauer parameters are reported in Table S2. The Mössbauer parameters of sample A agree with previously published values for LT-$Ni_3Sn_2$ [49]. For sample B, the values of the isomer shift obtained for the two Sn sites are lower than those reported for equi-stoichiometric $Ni_3Sn_4$ by about 0.05 mm.s⁻¹ and the quadrupole splitting of Sn(2) is higher by about 0.3 mm.s⁻¹ but they are close to the values of Ni-rich $Ni_{3.5}Sn_4$ [32]. The compositions of these two nickel-tin phases were evaluated from the average isomer shift by considering the linear correlation between average isomer shifts ($δ_{av}$) and atomic percentage of tin (Sn%) for different crystalline phases (Fig. S4). The compositions are $Ni_3Sn_2$ and $Ni_3.7Sn_4$ for samples A and B, respectively, in agreement with XRD and EPMA analysis. The Mössbauer spectrum of sample AB was fitted with four doublets corresponding to the two Sn crystallographic sites in $Ni_3Sn_2$ and $Ni_3Sn_4$, respectively (Fig. S3c), leading to Mössbauer parameters similar to those found for samples A and B except for the average isomer shift of $Ni_3Sn_2$ that is higher in sample AB than in sample A. The compositions obtained from $δ_{av}$ - Sn% linear correlation (Fig. S4) are $Ni_{3.7}Sn_2$ and $Ni_{2.6}Sn_2$. In the latter case, the composition differs from sample A but agrees with EPMA and XRD results. By neglecting the difference between the recoilless factors of the two nickel-tin phases in AB, their relative contributions to the overall spectrum give the relative amounts of Sn that are 56 at.% and 44 at.% for $Ni_3.7Sn_4$ and $Ni_{2.6}Sn_2$, respectively. This corresponds to about 50 wt.% for the two phases in close agreement with XRD analysis.

**Si/Ni-Sn/Al/C composites.** The Rietveld analysis of XRD patterns of ball-milled Si/Ni-Sn/Al/C composites are shown in Fig. 1 and structural data are gathered in Table 2.

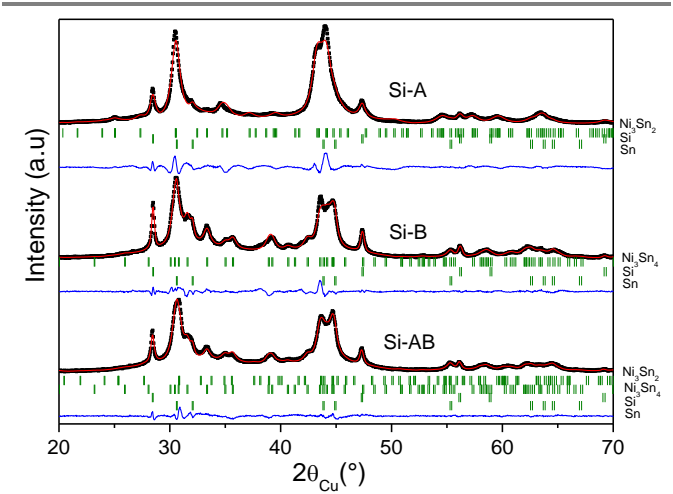

**Fig. 1.** XRD patterns and Rietveld analysis of composites Si/Ni-Sn/Al/C synthetized with different Ni-Sn intermetallic compounds: Si-A for $Ni_3Sn_2$; Si-B for $Ni_3Sn_4$ and Si-AB for $Ni_3Sn_2$-$Ni_3Sn_4$ mixture. Observed (dots), calculated (line) and difference (below) curves are shown. The vertical marks show the Bragg positions for all detected phases.

Main diffraction peaks were assigned to pure Si (peaks at 28.4°, 47.3° and 56.1°), $Ni_3Sn_2$ and $Ni_3Sn_4$. Carbon graphite transformed into disordered carbon during ball-milling. Minor contribution of β-Sn (< 5 wt.%) was detected as evidenced by a weak reflection around 31.5°. Minor tin formation indicates that side reaction between Si and Ni-Sn phases is of minor importance, *i.e.* both Si and intermetallic phases are essentially preserved during milling. This is attributed to the lubricating effect of carbon as PCA [38]. For the composite Si-A, the unit-cell volume of $Ni_3Sn_2$ is unchanged as compared to the pristine intermetallic (sample A). In contrast, for sample Si-B, the unit-cell volume of $Ni_3Sn_4$ phase, $V = 250.5$ Å$^3$, is significantly reduced relative to sample B, $V = 256.6$ Å$^3$. This is assigned to the extrusion of Ni from the $Ni_{3.6}Sn_4$ compound on milling. We hypothesize that the extruded Ni reacts with Si to form $NiSi_2$. Unfortunately, this phase is difficult to differentiate from pure Si by XRD due to severe peak overlapping [38]. According to the linear dependence between $Ni_3Sn_4$ cell volume and Ni-stoichiometry [43], the Ni content is reduced on milling from $Ni_{3.6}Sn_4$ to $Ni_{3.1}Sn_4$. For the composite Si-AB, Ni extrusion from the $Ni_3Sn_4$ phase is observed to the same extent. In addition, the unit cell volume of the $Ni_3Sn_2$ phase increases from 289.9 to 293 Å$^3$, which indicates a slight Ni-enrichment in the $Ni_3Sn_2$ phase from $Ni_{2.6}Sn_2$ to $Ni_{2.7}Sn_2$ [47]. The XRD patterns exhibit strong peak broadening which reflects the nanostructured state of the composites. The crystallite sizes of Ni-Sn intermetallics and silicon are comprised between 7 and 35 nm.

**Table 2.** Structural data of the Si-A, Si-B and Si-AB samples as determined from Rietveld analysis. Standard deviations referred to the last digit are given in parenthesis.

| Sample | Phase composition | S.G. | Unit-cell parameters | | | | | Amount (wt.%) | Crystallite size (nm) |
|---|---|---|---|---|---|---|---|---|---|
| | | | a, Å | b, Å | c, Å | β (°) | V, Å$^3$ | | |
| Si-A | $Ni_3Sn_2$ | Pnma | 7.112(6) | 5.166(2) | 8.207(7) | - | 301.6(4) | 76(6) | 7(2) |
| | Si | Fd$\bar{3}$m | 5.430(2) | - | - | - | 160.1(2) | 21(4) | 11(3) |
| | β-Sn | I4$_1$/amd | 5.840* | - | 3.180* | - | 108.4 | 3(1) | - |
| Si-AB | $Ni_{2.7}Sn_2$ | Pnma | 7.028(9) | 5.139(3) | 8.11(1) | - | 293.0(6) | 22(2) | 9(3) |
| | $Ni_{3.1}Sn_4$ | C2/m | 12.30(1) | 4.051(2) | 5.205(2) | 104.7(1) | 250.9(2) | 54(3) | 9(3) |
| | Si | Fd$\bar{3}$m | 5.435(2) | - | - | - | 160.5(2) | 20(2) | 22(7) |
| | β-Sn | I4$_1$/amd | 5.840* | - | 3.180* | - | 108.4 | 4(1) | - |
| Si-B | $Ni_{3.1}Sn_4$ | C2/m | 12.29(1) | 4.051(2) | 5.203(2) | 104.7(1) | 250.5(2) | 73(5) | 10(3) |
| | Si | Fd$\bar{3}$m | 5.427(2) | - | - | - | 159.9(2) | 23(3) | 35(9) |
| | β-Sn | I4$_1$/amd | 5.840* | - | 3.180* | - | 108.4 | 4(1) | - |

The Mössbauer spectra of the composites Si-A, Si-B and Si-AB are all formed by a more asymmetrical doublet than samples A, B and AB, respectively, suggesting the existence of Sn(0) similar to β-Sn as observed by XRD (Fig. 2). In addition, there is a shoulder close to 0 mm.s$^{-1}$ that reveals the existence of Sn(IV) oxides. The Mössbauer spectra of Si-A (Fig. 2a) and Si-B (Fig. 2b) were fitted with two doublets for reflecting the two Sn crystallographic sites of the intermetallic phases and Lorentzian curves for Sn(IV) and Sn(0). Another doublet at ~3 mm s$^{-1}$ was required to fit the Mössbauer experimental data of Si-A that can be assigned to Sn(II) oxides. The Mössbauer parameters are reported in Table 3. For the composite Si-A, the contribution of the doublets of intermetallic phase represents 90% of the overall spectrum and the values of the Mössbauer parameters are similar to those obtained for sample A. This confirms that ball milling process does not modify the stoichiometry of $Ni_3Sn_2$ as observed by XRD. The contributions to the spectrum of tin oxides represent 6% while that of β-Sn is 4%. For Si-B the contribution of the intermetallic is 92% of the overall spectrum while Sn(IV) oxides and β-Sn represent 3% and 5%, respectively. The value of the average isomer shift of the intermetallic (2.03 mm.s$^{-1}$) strongly differs from that obtained for sample B (1.93 mm.s$^{-1}$), which reflects change in the composition from $Ni_{3.7}Sn_4$ to $Ni_3Sn_4$. This confirms that Ni atoms are extruded from $Ni_{3.7}Sn_4$ during ball milling as detected by XRD.

As shown by XRD, the composite Si-AB contains two intermetallic phases $Ni_3Sn_2$ and $Ni_3Sn_4$. Thus, the Mössbauer spectrum was fitted with two doublets per intermetallic and single Lorentzian curves for β-Sn and Sn(IV) oxides. Due to strong differences between the expected $Ni_3Sn_2/Ni_3Sn_4$ weight ratio ($r = 1$) and the XRD Rietveld analysis ($r = 0.4$), the Mössbauer spectrum was fitted by considering different fixed values of r. The experimental data were adequately fitted with different values of $r$ in the range 0.6-0.8 but the most consistent results were obtained for $r = 0.7$ (Fig. 2c). This value is closer to the expected weight ratio, evidencing issues on the determination of phase content by XRD related to large peak broadening or even partial amorphization of the intermetallics resulting from ball milling. Mössbauer spectroscopy is more sensitive to local structure than XRD. The values of the Mössbauer parameters of $Ni_3Sn_4$ are similar to those of sample Si-B while the average isomer shift of $Ni_3Sn_2$ is higher to that of sample A by 0.04 mm.s$^{-1}$, leading to the composition $Ni_{2.6}Sn_2$ instead of $Ni_{2.8}Sn_2$ (Table 3). The same trend is observed by XRD. It is worth noting that the linewidth, Γ, is systematically higher for the composites compared to the intermetallics, in line with higher local disorder around Sn in the ball milled samples.

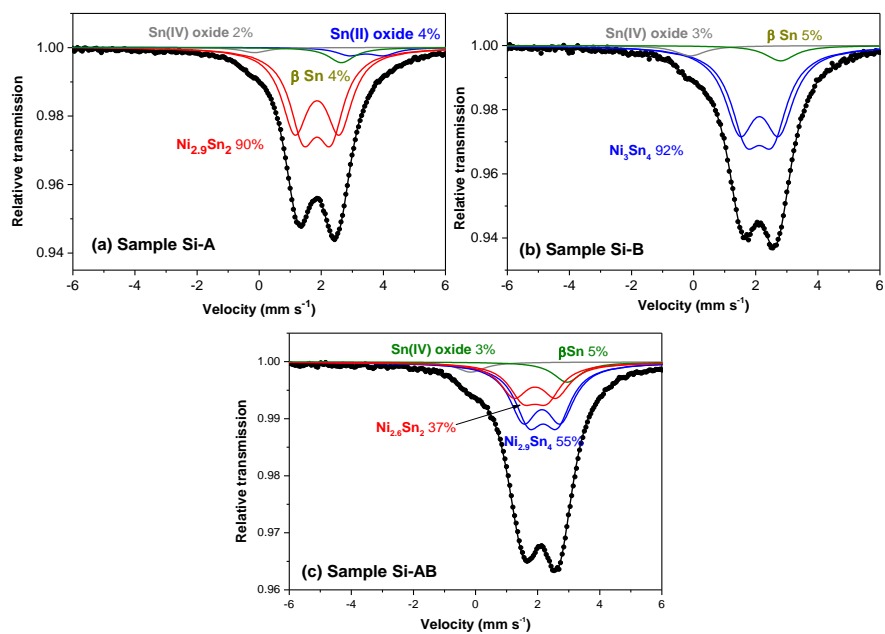

**Fig. 2**. $^{119}$Sn Mössbauer spectra of composites a) Si-A, b) Si-B and c) Si-AB at room temperature. The spectra were fitted with two doublets for each intermetallic, corresponding to the two Sn crystallographic sites, and components attributed to β-Sn and tin oxides. The subspectra corresponding to the different Sn crystallographic sites are shown with their relative contributions to the overall spectra.

**Table 3.** $^{119}$Sn Mössbauer parameters of composites Si-A, Si-B and Si-AB: isomer shift relative to BaSnO$_3$, $\delta$, quadrupole splitting, $\Delta$, linewidth, $\Gamma$, and relative spectral contribution, RSC, and the corresponding assigned phases. The uncertainties on Mössbauer parameters are lower than 0.05 mm.s$^{-1}$. The compositions of the intermetallics were evaluated from the average isomer shifts and linear correlation reported in Fig. S4.

| Sample | Phase composition | Mössbauer parameters | | | |
|---|---|---|---|---|---|
| | | $\delta$, mm.s$^{-1}$ | $\Delta$, mm.s$^{-1}$ | $\Gamma$, mm.s$^{-1}$ | RSC, % |
| Si-A | Ni$_{2.9}$Sn$_2$ | 1.76 | 1.43 | 1.02* | 45* |
| | | 1.76 | 0.89 | 1.02* | 45* |
| | β-Sn | 2.60 | - | 1.02* | 4 |
| | Sn(IV) | -0.20 | - | 1.02* | 2 |
| | Sn(II) | 3.3 | 1.1 | 1.02* | 4 |
| Si-AB | Ni$_{2.6}$Sn$_2$ | 1.80 | 1.37 | 1.08* | 29* |
| | | 1.80 | 0.79 | 1.08* | 29* |
| | Ni$_{2.9}$Sn$_4$ | 2.03 | 1.21 | 1.08* | 17** |
| | | 2.05 | 0.89 | 1.08* | 17** |
| | Sn(IV) | -0.25 | - | 1.08* | 3 |
| | β-Sn | 2.79 | - | 1.08* | 5 |
| Si-B | Ni$_3$Sn$_4$ | 2.02 | 1.26 | 1.16* | 46* |
| | | 2.01 | 0.86 | 1.16* | 46* |
| | Sn(IV) | -0.22 | - | 1.16* | 3 |
| | β-Sn | 2.70 | - | 1.16* | 5 |

*, **constrained to be equal

The influence of the intermetallic composition on the Si/Ni-Sn/Al/C composite microstructures is shown in Fig. 3. For the composite prepared with $Ni_3Sn_2$ (sample Si-A), the silicon nanoparticles (round black areas, ~ 150 nm in size) coexist with micrometric $Ni_3Sn_2$ domains (white areas). Phase identification is supported by EDX analysis [38]. Both Si and $Ni_3Sn_2$ particles are embedded into a grey matrix mainly attributed to aluminum and carbon. In contrast, for the composite prepared with $Ni_3Sn_4$ (sample Si-B), the constituents are homogeneously distributed at the nanoscale. Dark Si particles are embedded in a matrix with light-gray contrast attributed to a mixture of nanometric $Ni_3Sn_4$, aluminum and carbon. $Ni_3Sn_4$ domains are smaller than 50 nm as could be determined by TEM analysis (Fig. S5). For the composite prepared with the phase mixture $Ni_3Sn_2$-$Ni_3Sn_4$ (sample Si-AB), the constituents are also well-dispersed at the nanoscale though some micrometric $Ni_3Sn_2$ domains are detected.

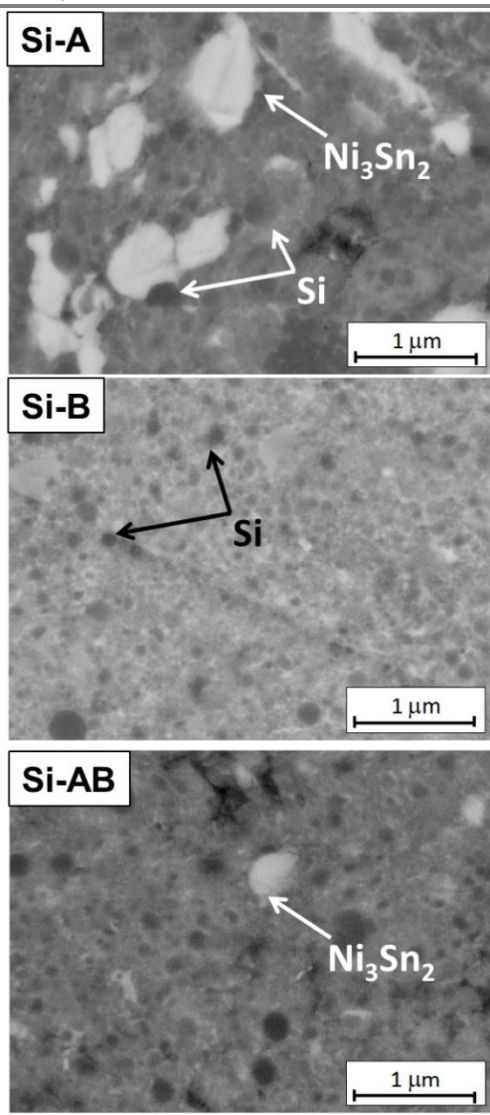

**Fig. 3.** SEM-BSE Cross-section images showing the microstructure of composites Si/Ni-Sn/Al/C synthesized with different Ni-Sn intermetallic compounds: Si-A for $Ni_3Sn_2$; Si-B for $Ni_3Sn_4$ and Si-AB for $Ni_3Sn_2$-$Ni_3Sn_4$ mixture.

## Electrochemical characterization

Fig. 4. shows the potential profiles of the first galvanostatic cycle for Si-A, Si-B and Si-AB. The three profiles are smooth, both in charge and discharge, showing a large sloping plateau between ~0.7 and 0 V. The composite Si-B shows the highest discharge and charge capacities with 869 and 756 mAh g$^{-1}$, respectively, corresponding to an initial coulombic efficiency, $\varepsilon_{c\text{-}ini}$, of 87%. The composites Si-AB and Si-A show a lower discharge/charge capacity with 730/578 mAh g$^{-1}$ and 684/481 mAh g$^{-1}$, and an initial coulombic efficiency decreasing with $Ni_3Sn_2$ content: 79.2 and 70.3 % for Si-AB and Si-A, respectively.

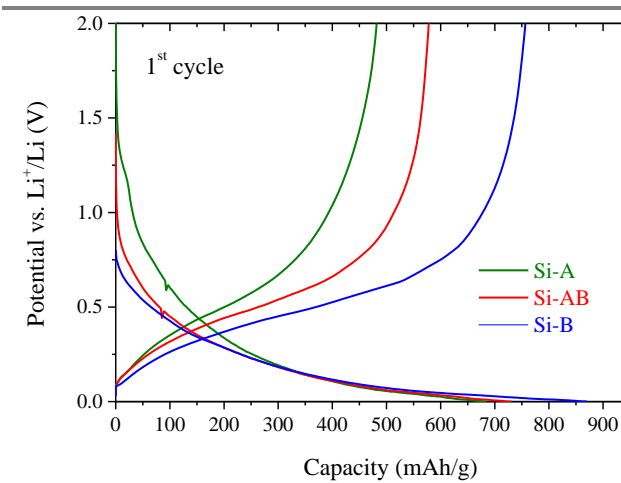

**Fig. 4.** Discharge/charge profiles of the first galvanostatic cycle for the three composite anodes Si-A, Si-B and Si-AB.

The evolution on cycling of the reversible capacity and coulombic efficiency for the three composites is shown in Fig. 5 and key data are gathered in Table 4. As a general trend, the initial reversible capacity increases with the amount of $Ni_3Sn_4$, being the highest for Si-B, $C_{ini}$ = 756 mAh g$^{-1}$. For all composites, the capacity decreases during the initial four cycles, then increases up to a maximum, $C_{max}$, at ~ 80 cycles and finally decreases slowly on further cycling. Interestingly, at the 200$^{th}$ cycle, the capacity of Si-AB, $C_{200}$ = 509 mAh g$^{-1}$ approaches that of Si-B, $C_{200}$ = 538 mAh g$^{-1}$. This convergence is driven by the lower capacity fade on cycling for the composite containing the biphasic matrix $Ni_3Sn_2$-$Ni_3Sn_4$ (0.043 %/cycle) as compared to that of single-phase $Ni_3Sn_4$ (0.094 %/cycle).

Fig. 5b shows the coulombic efficiency of the different composites. The efficiency on prolonged cycling improves with $Ni_3Sn_2$ amount. Thus, the highest average coulombic efficiency for cycles fourth to 200, $\varepsilon_{c4\text{-}200}$, is obtained for Si-A, 99.7% (Table 4). Strikingly, this trend differs with the evolution of the efficiency at the first cycle, $\varepsilon_{c\text{-}ini}$, which, as reported before, decreases with $Ni_3Sn_2$ amount. For testing the rate capability on this type of materials, the charge capacity of the Si-B composite as a function of kinetic regime was determined (Fig. S6). The charge capacity is stable up to 10 C.

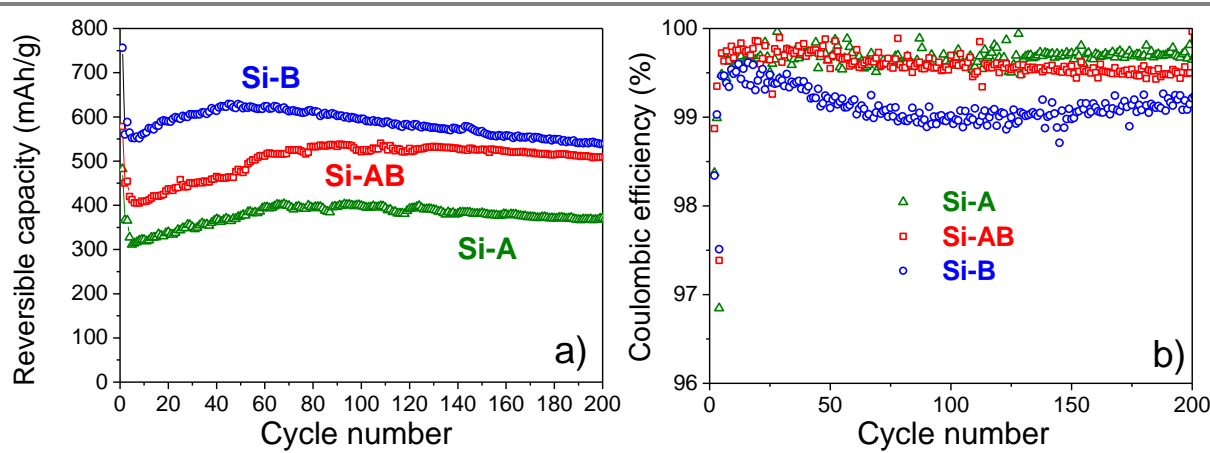

**Fig. 5.** a) Specific reversible capacity (charge) and b) coulombic efficiency of the Si/Ni-Sn/Al/C composites synthesized with different Ni-Sn intermetallic compounds: Si-A for $Ni_3Sn_2$; Si-B for $Ni_3Sn_4$ and Si-AB for $Ni_3Sn_2$-$Ni_3Sn_4$ mixture.

**Table 4.** Reversible specific capacities (charge) and coulombic efficiencies of the Si/Ni-Sn/Al/C composites: initial capacity ($C_{ini}$), maximum capacity ($C_{max}$), capacity at 200 cycles ($C_{200}$), average capacity between 4 and 200 cycles ($C_{4-200}$), capacity loss per cycle between $C_{max}$ and $C_{200}$ ($C_{loss}$), initial efficiency ($\varepsilon_{c-ini}$) and average efficiency between 4 and 200 cycles ($\varepsilon_{c,4-200}$)

| Sample | Nominal Composition | $C_{ini}$ (mAh g⁻¹) | $C_{max}$ (mAh g⁻¹) | $C_{200}$ (mAh g⁻¹) | $C_{4-200}$ (mAh g⁻¹) | $C_{loss}$ (%/cycle) | $\varepsilon_{c-ini}$ (%) | $\varepsilon_{c,4-200}$ (%) |
|---|---|---|---|---|---|---|---|---|
| Si-A | Si/$Ni_3Sn_2$/Al/C | 481 | 402 | 370 | 375 | 0.060 | 70.3 | 99.7 |
| Si-AB | Si/$Ni_3Sn_2$+$Ni_3Sn_4$/Al/C | 578 | 537 | 509 | 502 | 0.043 | 79.2 | 99.6 |
| Si-B | Si/$Ni_3Sn_4$/Si/Al/C | 756 | 628 | 538 | 584 | 0.094 | 87.0 | 99.1 |

## Discussion

Electrochemical results show that both the reversible capacity and coulombic efficiency of composites Si/Ni-Sn/Al/C increase with $Ni_3Sn_4$ amount at the first cycles but, on long-term cycling, capacity decay and coulombic efficiency improves with $Ni_3Sn_2$ amount (Fig. 5a). Indeed, as a compromise, the composite Si-AB made with the bi-phasic matrix $Ni_3Sn_2$/$Ni_3Sn_4$ provides the best electrochemical performances in terms of average capacity and coulombic efficiency. A better understanding of this major result could be gained at the light of the microstructural characterization of the composites (*cf.* section on Chemical and microstructural characterization) and after considering the reactivity of Ni-Sn intermetallics towards lithium. However, as for the latter, no flat plateau potentials are detected in discharge/charge profiles (Fig. 5) which makes difficult to identify reactions between the composite counterparts and lithium. Smooth profiles, which are observed all over 200 cycles (see Fig. S7 in supplementary information), result from the combination of multiple and poorly-defined potentials for the reaction between lithium and composite counterparts, namely Ni-Sn intermetallics, disordered carbon and nanostructured silicon [16,32,50–52].

To overcome this limitation, analysis of Differential Capacity Plots (DCPs) has been performed to identify electrochemical charge/discharge reactions and track their evolution during cycling. Fig. 6 shows the DCPs for the different Si/Ni-Sn/Al/C composites at cycles 2, 80 and 200 corresponding to composite activation, maximum capacity and end-of-cycling states. For Si-B, at the 2nd lithiation (Fig. 6a), three reduction peaks are observed in the cathodic branch. The peak at 0.17 V is attributed to the reaction potential of amorphous Si with lithium to form $Li_xSi$ alloys of approximate composition LiSi and $Li_7Si_3$ [45,53]. Silicon amorphisation is known to occur during the first lithiation of crystalline Si anodes [39,45]. The two other reduction peaks at 0.31 and 0.62 V are attributed to the formation of $Li_xSn$ alloys: $Li_7Sn_2$ and LiSn, respectively [34,54,55]. On delithiation process, multiple oxidation peaks are detected and are attributed to the decomposition of $Li_7Si_3$ at 0.51 V and different $Li_xSn$ alloys: $Li_7Sn_2$ (or $Li_{22}Sn_5$), $Li_5Sn_2$, LiSn and $Li_2Sn_5$ at 0.47, 0.58, 0.74 and 0.81 V, respectively [53,56,57]. The two phases $Li_7Sn_3$ and $Li_{13}Sn_5$ have compositions close to $Li_5Sn_2$ and could be involved in the delithiation process at 0.58 V. For both composites Si-AB and Si-A, poorly defined cathodic peaks are found at the 2nd lithiation, but DCPs exhibit significant features in delithiation. For Si-AB, two oxidation peaks are clearly identified at 0.47 and 0.58 V attributed to decomposition of $Li_7Sn_2$ (or $Li_{22}Sn_5$) and $Li_5Sn_2$ alloys, respectively. For Si-A, a unique broad oxidation peak is observed at 0.51V assigned to the decomposition of $Li_7Si_3$ [20,53]. Interestingly, no signal related to decomposition of Li$_y$Sn alloys is detected for the composite Si-A, suggesting that Ni$_3$Sn$_2$ does not react with lithium.

The DCPs at the maximum capacity, cycle 80 (Fig. 6b), show significant differences as compared to those for the 2$^{nd}$ cycle. In the anodic branch, a new oxidation peak appears at 0.35 V for all composites. It is attributed to the decomposition of a Li-rich Li$_x$Si alloy with approximate composition of Li$_{3.16}$Si[20,53], i.e. with higher Li-content than for Li$_7$Si$_3$ (oxidation peak at 0.51 V). Therefore, as compared to the 2$^{nd}$ cycle, silicon nanoparticles reach higher lithium content. Such a gradual and deeper lithiation of Si can explain the capacity maximum observed in Fig. 5, at cycle ~ 80, for all composites. In contrast, oxidation peaks at 0.47 and 0.58 V related to decomposition of Li-rich Li$_y$Sn alloys are not detected at cycle 80, suggesting that only Li-poor LiSn and Li$_2$Sn$_5$ alloys are formed for Si-B and Si-AB. As concerns cycle 200 (Fig. 6c), very slight differences are found compared to cycle 80 (Fig. 6b). No new peaks are detected, though the peak position related to the formation of Li$_y$Sn alloys slightly shifts to lower voltages with Ni$_3$Sn$_4$ concentration: 0.60 and 0.55 V for the composites Si-AB and Si-B, respectively. Moreover, for the composite Si-B, the intensity of the oxidation peaks at high potential, 0.66 and 0.81 V, assigned to the decomposition of Li-poor Li$_y$Sn alloys tends to decrease, suggesting that the capacity decay for this composite is mainly due to degradation of the electrochemical properties of the Ni$_3$Sn$_4$ counterpart. Besides identification and evolution on cycling of electrochemical reactions, DCPs clearly show, as a general trend, that the area of both anodic and cathodic branches decreases with Ni$_3$Sn$_2$ amount (Fig. 6). This concurs with the overall decrease in reversible capacity along the series Si-A, Si-AB and Si-B (Fig.7a). Moreover, as mentioned before, no reduction/oxidation peaks related to the formation/decomposition of Li$_x$Sn alloys is detected in DCPs for the composite Si-A whatever the galvanostatic cycle is. Since the Si content is constant in all composites, (19 wt.%), it can be inferred that Ni$_3$Sn$_4$ reacts with Li, whereas Ni$_3$Sn$_2$ does not. To verify this hypothesis, $^{119}$Sn Mössbauer spectroscopy studies have been undertaken for the composites Si-A and Si-AB at the end of first discharge (Fig. 7). The Mössbauer spectrum of Si-A (Fig. 7a) shows that the major phase is Ni$_3$Sn$_2$, with a contribution close to that of the pristine material, and an unresolved doublet that could be attributed to Li$_7$Sn$_2$ (Table 5). The latter phase arises from the lithiation of secondary β-Sn and tin oxides observed in Si-A (Fig. 2a). This clearly indicates that Ni$_3$Sn$_2$ is electrochemically inactive. In contrast, the Mössbauer spectrum of Si-AB (Fig. 7b) reflects the existence of Ni$_3$Sn$_2$ and high amount of Li-rich Li$_x$Sn alloys (Table 5). The isomer shift of the latter Li$_x$Sn phase (1.75 mm.s$^{-1}$) is close to that of Li$_{22}$Sn$_5$, which suggests a deeper lithiation than sample Si-A ending with Li$_7$Sn$_2$. This means that Ni$_3$Sn$_4$ reacts with lithium to form Li$_x$Sn alloys while Ni$_3$Sn$_2$ is, again, electrochemically inactive.

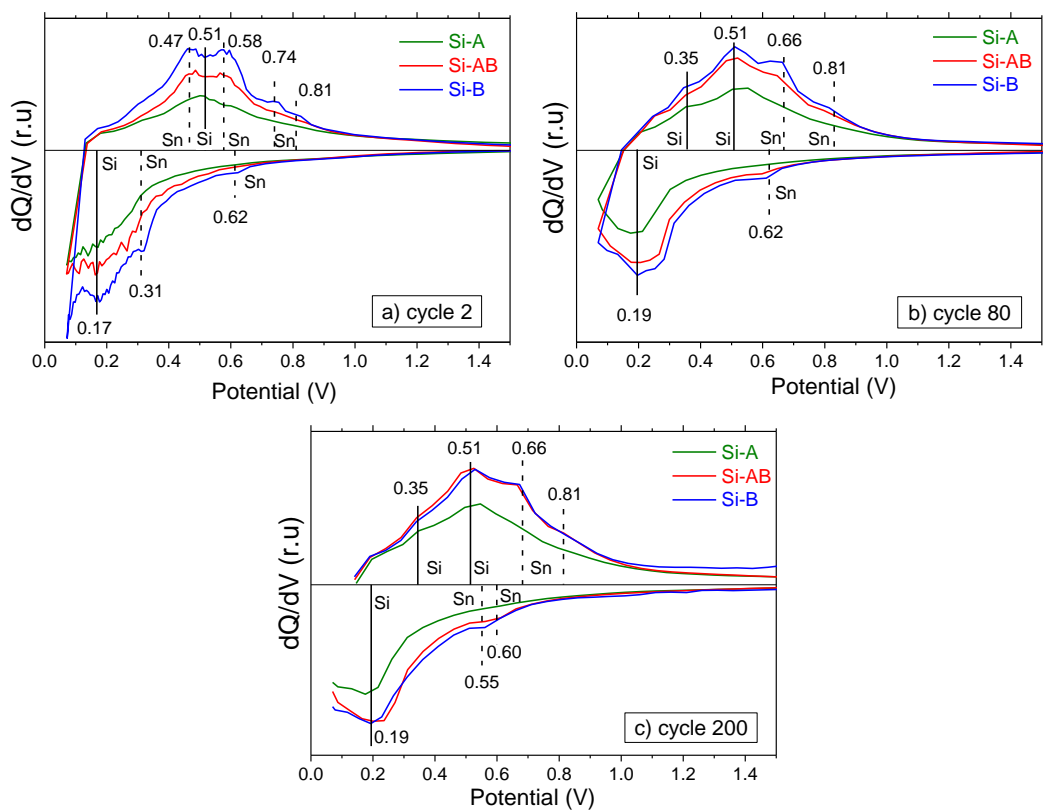

**Fig. 6.** d$Q$/d$V$ dependence as a function of voltage for the composites Si-A, Si-AB and Si-B at cycles a) 2, b) 80 c) 200. Peak assignation to formation/decomposition of different Li$_x$Si and Li$_y$Sn alloys is denoted by solid and dashed vertical lines, respectively.

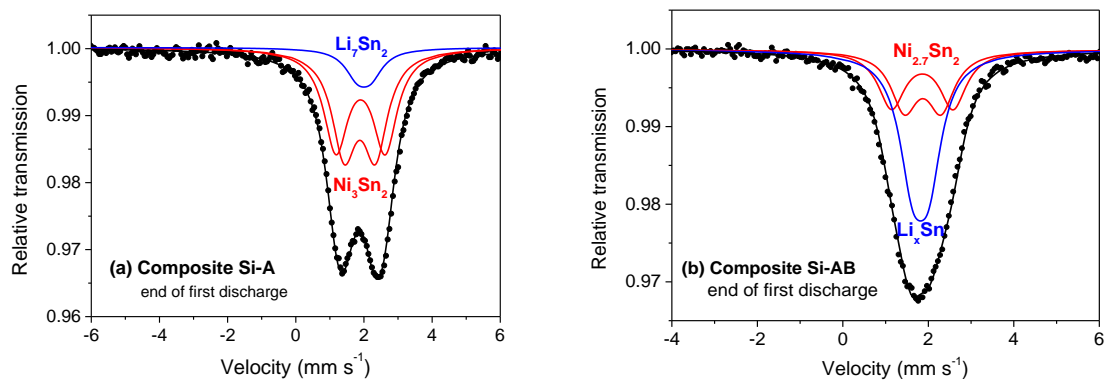

**Fig. 7.** $^{119}$Sn Mössbauer spectra of composites a) Si-A and b) Si-AB at the end of the first discharge obtained at room temperature. The spectra were fitted with two doublets corresponding to the two Sn crystallographic sites in LT-Ni$_3$Sn$_2$ and a doublet attributed to Li$_x$Sn alloys. The subspectra corresponding to the different Sn crystallographic sites are shown with their relative contributions to the overall spectra.

**Table 5.** $^{119}$Sn Mössbauer parameters of Si-A and Si-AB electrode materials at the end of discharge: isomer shift relative to BaSnO$_3$, $\delta$, quadrupole splitting, $\Delta$, linewidth, $\Gamma$, and Relative Spectral Contribution, RSC, and the corresponding assigned phases. The uncertainties on Mössbauer parameters are lower than 0.05 mm s$^{-1}$. The compositions of the intermetallics were evaluated from the average isomer shifts and linear correlation reported in Fig. S4.

| Sample | Phases | Mössbauer parameters | | | |
|---|---|---|---|---|---|
| | | $\delta$, mm.s$^{-1}$ | $\Delta$, mm.s$^{-1}$ | $\Gamma$, mm.s$^{-1}$ | RSC, % |
| A | Ni$_{2.8}$Sn$_2$ | 1.78 | 1.44 | 0.86* | 44* |
| | | 1.76 | 0.92 | 0.86* | 44* |
| | Li$_7$Sn$_2$ | 1.89 | 0.40 | 0.86* | 12 |
| AB | Ni$_{2.7}$Sn$_2$ | 1.78 | 1.40 | 0.80* | 24* |
| | | 1.79 | 0.86 | 0.80* | 24* |
| | Li$_x$Sn | 1.75 | 0.40 | 0.80* | 52 |
| * constrained to be equal | | | | | |

To summarize, Si and Ni$_3$Sn$_4$ provide the most substantial contributions to the overall composite capacity. DCPs show that, at the deepest lithiation state, these constituents form Li$_{3.16}$Si and Li$_7$Sn$_2$ alloys, respectively. From their phase amount and lithiation capacities, the maximum overall capacity of the composites was estimated and compared with the initial measured capacity (see Table S3 in supplementary information). Whatever the composite, the initial specific capacity of the electrode is roughly 80% of the maximum overall capacity. This concurs with the fact that electrodes are not fully activated at the first cycle. For composite Si-B, 60 % of the capacity is provided by Si, whereas 40% is delivered by Ni$_3$Sn$_4$.

It is widely reported that the intermetallic compound Ni$_3$Sn$_4$ reacts with lithium according to the conversion reaction [39,49]

$$Ni_3Sn_4 + 14\,Li^+ + 14e^- \rightarrow 2Li_7Sn_2 + 3Ni \qquad R1$$

The delithiation reaction occurs through the formation of Li-poor Li$_x$Sn alloys before reformation of the intermetallic compound Ni$_3$Sn$_4$. The lithiation reaction only occurs significantly when the intermetallic compound is nanostructured [32]. CV features of reversible Ni$_3$Sn$_4$ lithiation are very close to those of β-Sn [32,34,55] in agreement with DCP plots (Fig. 6). In this work, the fact that Ni$_3$Sn$_4$ is highly nanostructured forming sub-micrometric domains (Fig. 3) allows its reversible reaction with lithium.

As for the reactivity of Ni$_3$Sn$_2$, the interpretation is more intricate. For long time, this compound has been considered having very low reactivity against lithium [58,59]. However, Yi et al. have recently reported that this compound significantly reacts with lithium according to the reaction scheme [52]:

$$Ni_3Sn_2 + 8.8\,Li^+ + 8.8e^- \rightarrow 2Li_{4.4}Sn + 3\,Ni \qquad R2$$

However, our DCPs (Fig. 6) and Mössbauer analysis (Fig. 7) demonstrate that Ni$_3$Sn$_2$ is inactive in this work. The reason for this apparent contradiction can be found in the microstructural analysis (Fig. 3), which shows that Ni$_3$Sn$_2$ domains are micrometric. Indeed, all reports claiming reactivity of Ni$_3$Sn$_2$ against lithium refer to different nanometric materials: thin films [58,60], nanoparticles [61] and nanocomposites [50]. Interestingly, though micrometric Ni$_3$Sn$_2$ domains are inactive, some reactivity at the nanoscale may facilitate Li-diffusion along Ni$_3$Sn$_2$/Si and Ni$_3$Sn$_2$/Ni$_3$Sn$_4$ interfaces, behaving as a good solid-state ionic conductor. A similar role has been assigned to Ti$_6$Ni$_4$Si$_7$ for the fast lithiation of Si/Ti$_6$Ni$_4$Si$_7$ composites.[24]

The size of both phase domains, Ni$_3$Sn$_2$ and Ni$_3$Sn$_4$, in the buffering matrix plays a key role not only on their intrinsic reactivity but also on the coulombic efficiency of the whole composite. It is worth noting that the worst initial coulombic efficiency is found for Si-A composite, $\varepsilon_{c\text{-ini}}$ = 70.3%. Since Ni$_3$Sn$_2$ is inactive and micrometric, such a low coulombic efficiency cannot be attributed to irreversible lithiation of this phase, neither to significant electrolyte decomposition at its poorly developed surface. Consequently, it is assigned to the Si-counterpart which volume swelling is not efficiently buffered by the micrometric intermetallic matrix. In contrast, Ni$_3$Sn$_4$ which is active and forms nanometric domains, efficiently buffers and protects Si nanoparticles at short cycling offering the best initial coulombic efficiency, $\varepsilon_{c\text{-ini}}$ = 87.0%. Unfortunately, for long-term cycling, the coulombic efficiency of the Si-B composite is unsatisfactory, $\varepsilon_{c,4\text{-}200}$ = 99.1%, which may be linked either to high surface area or to the significant volume expansion of the electrode due to reactivity of both Si and $Ni_3Sn_4$ phases. Thus, the best compromise for Si/Ni-Sn/Al/C composites in terms of coulombic efficiency, at short- and long-term cycling, is found for the bi-phasic matrix $Ni_3Sn_2/Ni_3Sn_4$ in which inactive $Ni_3Sn_2$ minimizes electrode swelling and limits electrolyte degradation whereas $Ni_3Sn_4$ efficiently buffers volume changes of silicon nanoparticles.

## Conclusion

To bring Si-based anodes to the market, advanced 3D materials at affordable cost obtained by high-throughput synthetic routes should be implemented. These materials must provide reversible capacities significantly higher than that of current graphite electrodes, with coulombic efficiency exceeding 99.9 % and cycle-life exceeding hundreds of cycles. Composite materials made of Si/Ni-Sn/Al/C produced by classical milling methods are promising materials to fulfil these goals.

The composite materials studied in this work consist of Si nanoparticles embedded in multi-phase matrix that contains aluminum, carbon and Ni-Sn intermetallics. To reach long-term cycling, the multi-phase matrix must efficiently accommodate Si volume changes, which can be achieved when the matrix counterparts are intimately mixed at the nanoscale. In this research, we have investigated the effect of three different Ni-Sn counterparts ($Ni_3Sn_4$, $Ni_3Sn_2$ and their mixture) on the buffering efficiency of the matrix. Best results as regards long-term cycling are obtained for the bi-phasic mixture in which $Ni_3Sn_4$ is electrochemically active whereas $Ni_3Sn_2$ not. Additionally, compared to previous works on Si/Ni-Sn/Al/C composites, the partial replacement of $Ni_3Sn_4$ by $Ni_3Sn_2$ significantly reduces the Sn content in the material and therefore its cost. Nonetheless, the average coulombic efficiency of the Si/$Ni_3Sn_2$-$Ni_3Sn_4$/Al/C system, though rather high 99.6 %, is still unsatisfactory for practical applications and the irreversibility at the first cycle, 20.8 %, needs to be further corrected. A higher nanostructuration of the intermetallic counterpart $Ni_3Sn_2$, for instance through a pre-milling treatment, and further research work on composite coating treatment and electrode formulation are foreseen as promising solutions to overcome these remaining issues.

## Conflicts of interest

There are no conflicts to declare.

## Acknowledgments

The authors are grateful to Remy Pires for SEM analysis, Eric Leroy for EPMA analysis and to the French Research Agency ANR (project NEWMASTE, n°ANR-13-PRGE-0010) for financial support.